\documentclass[final,3p,times]{elsarticle}
\makeatletter
\def\ps@pprintTitle{%
  \let\@oddhead\@empty
  \let\@evenhead\@empty
  \def\@oddfoot{\reset@font\hfil\thepage\hfil}
  \let\@evenfoot\@oddfoot
}
\makeatother
\usepackage{graphicx,epsfig}
\usepackage{graphicx,subfigure}
\usepackage{amsmath}
\usepackage{amssymb}
\usepackage{amsthm}
\usepackage{mathrsfs}
\usepackage{amsfonts}
\usepackage{amsfonts}
\usepackage{setspace}
\usepackage{mathtools}
\usepackage{makecell}
\usepackage{graphicx,epsfig}
\usepackage{graphicx,subfigure}
\usepackage[colorlinks = true,
            linkcolor = blue,
            urlcolor  = blue,
            citecolor = blue,
            anchorcolor = blue]{hyperref}
\usepackage{caption}
 \makeatletter
\input epsf
\newcounter{saveeqn}

\begin{document}

\begin{frontmatter}

\title{Scattering of oblique waves by permeable vertical flexible membrane wave barriers}

\author[mymainaddress]{S. Koley\corref{mycorrespondingauthor}}
\cortext[mycorrespondingauthor]{Corresponding author}
\ead{santanukoley1989@gmail.com, santanukoley1989@yahoo.com}
\author[mymainaddress]{T. Sahoo}
\ead{tsahoo1967@gmail.com, tsahoo@naval.iitkgp.ernet.in}
\address[mymainaddress]{Department of Ocean Engineering and Naval Architecture,\\
Indian Institute of Technology, Kharagpur -721 302, India}

\begin{abstract}

The interaction of obliquely incident surface gravity waves with a vertical flexible permeable membrane wave barrier is investigated in the context of
three-dimensional linear wave-structure interaction theory. A general formulation for wave interaction with permeable submerged vertical membrane is given. The analytic solution of the physical problem is obtained by using  eigenfunction expansion method, and boundary element method has been used to get the numerical solution. In the boundary element method, since the boundary condition on the membrane is not known in advance, membrane motions and velocity potentials are solved simultaneously. From the general formulation of the submerged membrane barrier, the performance of bottom-standing, surface-piercing and fully extended membrane wave barriers are analyzed for various wave and structural parameters. It is found that the efficiency of the submerged,  surface-piercing and bottom-standing membrane wave barriers can be enhanced in waves for certain design conditions. From the analysis of various membrane configurations and parameters, it can be concluded that permeable membrane wave barrier can function as a very effective breakwater if it is properly designed.

\end{abstract}


\end{frontmatter}

\section{Introduction}
\label{sec:1}
In recent decades, flexible porous breakwaters have been considered as better alternatives to the conventional fixed rigid breakwaters for providing protection from wave attack at locations where protection is required on temporary basis. These types of structures are more suitable where bottom soil foundation is poor as these types of structures do not required proper foundation or strong supports. Moreover, flexible permeable barriers are cost-effective, quickly deployable, lightweight, portable, reusable and environmental friendly. Due to the porosity, these structures can dissipate wave energy at a higher rate which in turn reduces wave forces on the structure. Moreover, a characteristic of flexibility is usually included in these temporary barriers in order to minimize the wave impact on them. In addition, partial flexible barriers allow the free water circulation, transportation of sediment and safe passage of ocean current. Often, permeable and flexible structures are used to reduce wave resonance inside the harbor along with both the reflected and the transmitted wave heights during wave scattering.

\cite{williams1991flexible} have investigated the performance of a flexible, floating beam-like structure anchored to the seabed and having a small buoyancy chamber at the top of structure. They have solved the physical problem numerically by using the boundary integral equation method and validated the results by carried out small-scale physical model tests. \cite{kim1996flexible} studied the interaction of water waves with a vertical flexible membrane in the context of two-dimensional linear water wave theory. \cite{cho1997performance} have studied the interaction of oblique incident waves with a tensioned vertical flexible membrane hinged at the sea floor and attached to a rigid cylindrical buoy at its top. By using Darcy's fine-pore model, \cite{cho2000interactions} studied the interaction of monochromatic incident waves with a horizontal flexible porous  membrane in the context of 2D linear hydroelastic theory. To restore the wetlands habitat, \cite{williams2003flexible} proposed to use flexible porous wave barrier to protect cordgrass seedlings form wave action during the initial stage of growth following plating. \cite{kumar2006wave} have analyzed the performance of a vertical flexible porous breakwater in two-layer fluid of finite depth. Further, \cite{kumar2007wave} have carried out an analysis to investigate the scattering of water waves by a vertical flexible porous membrane pinned both at the free surface and the sea bed in a two-layer fluid of finite depth. \cite{karmakar2013scattering} used least square approximation method to study the scattering of surface gravity waves by multiple surface-piercing flexible permeable membrane
wave barriers. Using system of Fredholm integral equation approach, \cite{koley2015oblique} and \cite{kaligatla2015trapping} studied the interaction of surface gravity waves by a floating flexible porous plate in water of finite and infinite depths. Recently, using the same approach as used in  \cite{koley2015oblique}, \cite{koley2016membrane} analyzed the hydroelastic response of a floating horizontal flexible porous membrane.

In the present study, scattering of obliquely incident surface gravity waves with a submerged permeable vertical flexible membrane barrier in water of finite depth is investigated in the context of three-dimensional linear water wave theory. As special cases of the submerged barrier, the effectiveness of the bottom-standing as well as surface-piercing and complete membrane barrier are analyzed. The solution of the boundary value problem associated with each barrier configuration is obtained (i) analytically by using eigenfunction expansion method and (ii) numerically by using boundary element method. Finally, both the results are compared for accuracy and convergence. The boundary element method is developed based on discrete membrane dynamic model and simple-source distribution over the entire fluid boundaries. To understand the efficiency of the proposed system, the reflection and transmission coefficients, the wave forces acting on the structure and the structural displacements of the membrane have been plotted and analyzed for various values of wave and structural parameters.

\section{Mathematical formulation}
\label{sec:2}
\begin{figure}[ht!]
\begin{center}
\subfigure[]{\label{fig:1a}\includegraphics[width=8.4cm, height=4.5cm]{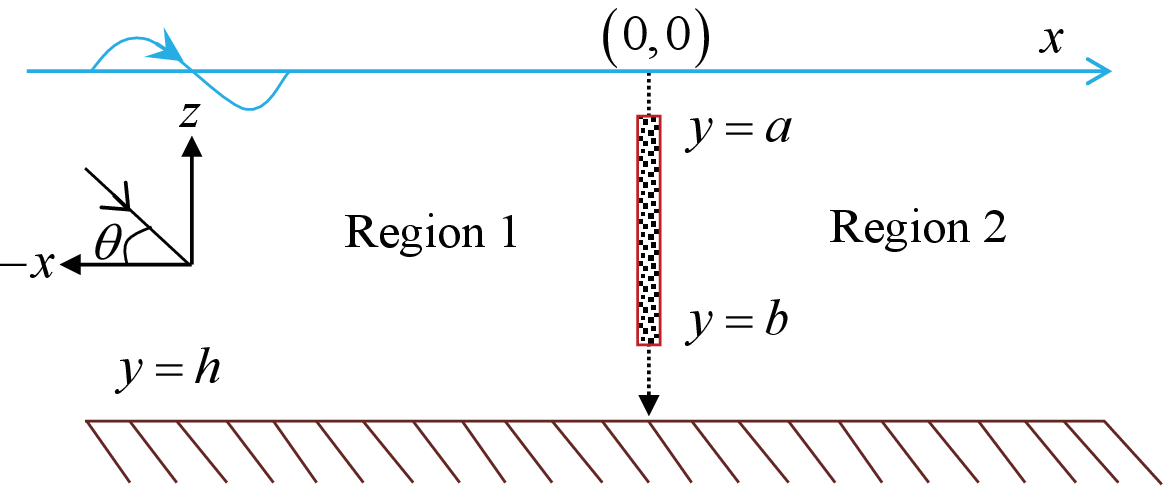}}
\subfigure[]{\label{fig:1b}\includegraphics[width=8.4cm, height=4.5cm]{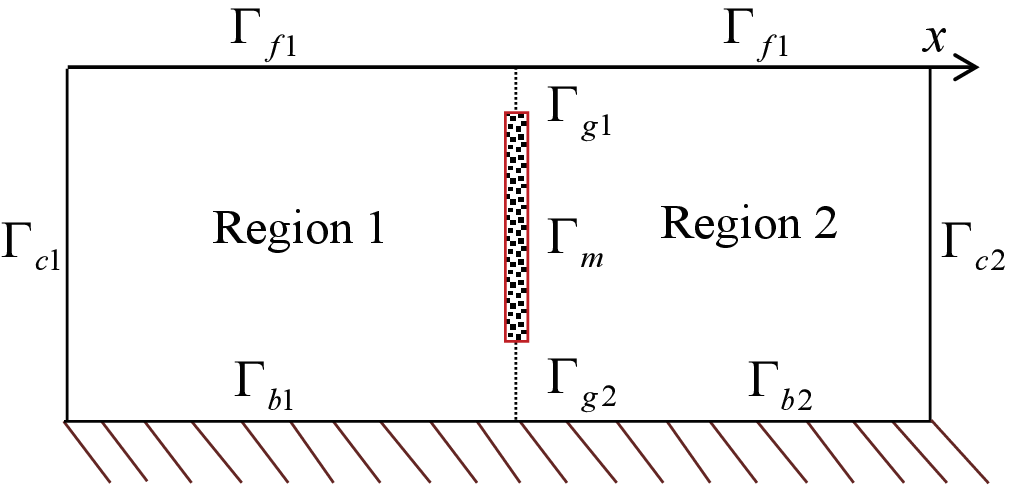}}\vspace{-0.5cm}
\caption{Schematic diagrams of wave scattering by permeable flexible membrane wave barrier.}\label{fig:1}
\end{center}
\end{figure}

In the present manuscript, oblique wave scattering by a submerged flexible porous membrane is studied in water of finite depth under the assumption of small amplitude water wave theory
and structural responses. The physical problems are analyzed in the three-dimensional Cartesian coordinate system with the positive $y$-axis being vertically downwards and the
horizontal plane $y=0$ represents undisturbed free surface. The fluid domain is infinitely extended in the $x$-$z$ horizontal direction as $-\infty<x, z<\infty$ except the flexible porous
membrane as in Fig. \ref{fig:1}a. A thin vertical flexible porous membrane occupies the region $x=0, a<y<b, -\infty<z<\infty$ in the fluid domain. For notational convenience,
$L_m=\left(a, b\right)$ refers to the membrane segment and $L_g=\left(0, a\right) \cup \left(b, h\right)$ refers to the gap region. The fluid is modeled using the Airy's water wave theory,
and the motion of the fluid is assumed to be of simple harmonic in time with the angular frequency $\omega$. Further, it is assumed that surface waves are incident upon the vertical membrane by making an angle $\theta$ with the $x$- axis. These assumptions ensure the existence of velocity potential $\Phi \left( x,y,z,t\right)$ and is of the form $\Phi \left( x,y,z,t\right) = \mathrm{Re}\left\{ \phi_j (x, y) e^{\mathrm{i}\left(\beta_0 z - \omega t\right)}\right\}$ with $\beta_0=k_0 \sin \theta$ being the $z$-component of the plane progressive wave incident
upon the membrane and the subscripts $j=1, 2$ referring to the fluid domains 1 and 2 as shown in Fig. \ref{fig:1}a. In the $j^{\mathrm{th}}$ fluid region, the spatial velocity
potential $\phi_j$ satisfies the partial differential equation
\begin{equation} \label{eq:1}
\left(\frac{\partial^{2}}{\partial x^{2}} +
\frac{\partial^{2}}{\partial y^{2}}-\beta_0^2\right)\phi_j = 0.
\end{equation}
The free surface boundary condition is given by
\begin{equation} \label{eq:2}
\frac{\partial\phi_j}{\partial y} + K \phi_j=0,~~\mbox{on}~~y=0,
\end{equation}
where $K=\omega^2/g$ with $g$ being the acceleration due to gravity. The bottom boundary condition
is given by
\begin{equation}\label{eq:3}
\displaystyle \frac{\partial \phi_j}{\partial y}=0,~~\mbox{on}~~y=h,
\end{equation}
The membrane barrier is modeled as string of uniform mass density $m_m$ acting under uniform tension with both the ends being fixed. The motion of the membrane barrier is assumed to be uniform in the longitudinal direction and the barrier is deflected horizontally with a displacement of the form $\zeta (y,z,t) = \mbox{Re}\left\{\chi (y) e^{\mathrm{i}(\beta_0 z-\omega t)}\right\}$ with $\chi (y)$ being the complex deflection amplitude and is assumed to be small compared to the water depth. The equation of motion of the membrane deflection $\chi (y)$, acted upon by the dynamic pressure is given by
\begin{equation} \label{eq:4}
\overline{T} \left(\frac{d^2}{dy^2}-\beta_0^2\right) \chi + m_m \omega^2 \chi=-\mathrm{i}\rho \omega \left[\phi_1(0, y) - \phi_2(0, y)\right],~~\mbox{for}~~x=0, y\in L_{m},
\end{equation}
where $\overline{T}$ is the membrane tension, $\rho$ is the density of the water and $m_m=\rho_m d_m$ is the uniform mass per unit length of the membrane with thickness $d_m$ and density $\rho_m$. Further, the boundary condition on the flexible porous membrane with porous-effect parameter $G_0$ (see \cite{yu1995diffraction}) is given by
\begin{equation} \label{eq:5}
\frac{\partial \phi_j}{\partial x}=\mathrm{i} k_0 G_0 \left[\phi_1(0, y) - \phi_2(0, y)\right] - \mathrm{i} \omega \chi,~~\mbox{for}~~x=0, y\in L_{m}.
\end{equation}
It may be noted that in Eq. (\ref{eq:5}), $k_0$ is the wave number associated with the plane progressive wave satisfying the dispersion relation $\omega^2=gk_0\tanh k_0h$.
Assuming that the membrane is fixed at both the ends, the vanishing of the membrane deflection yields
\begin{equation}\label{eq:6}
\chi=0, \quad \mbox{at}\quad y=a, b
\end{equation}
Further, the continuity of pressure and velocity in the gap region yield
\begin{equation}\label{eq:7}
\phi_1=\phi_2, \quad \frac{\partial \phi_1}{\partial x}=\frac{\partial \phi_2}{\partial x},~~\mbox{for}~~x=0, y\in L_g.
\end{equation}
Finally, the far-field boundary conditions are given by
\begin{equation}\label{eq:8}
\left\{\begin{array}{l}
\phi_1\left(x, y\right) \rightarrow \phi_0 \left(x, y\right) + R \phi_0 \left(-x, y\right) \quad \mbox{as} x \rightarrow -\infty, \vspace{0.2cm}
\\
\phi_2\left(x, y\right) \rightarrow T \phi_0 \left(x, y\right) \quad \mbox{as} x \rightarrow \infty,
\end{array}\right.
\end{equation}
where $R$ and $T$ are the complex coefficients associated with the reflected and transmitted waves. In Eq. (\ref{eq:8}), $\phi_0 (x,y)$ is the incident wave potential and is given by
\begin{equation}\label{eq:8a}
\displaystyle \phi_0 \left(x,y\right)=-\frac{\mathrm{i}gH}{2\omega} \frac{\cosh\left(k_0(h-y)\right)}{\cosh k_0h} e^{\mathrm{i}\alpha_0 x},
\end{equation}
with $\alpha_0=k_0 \cos \theta$ and $H$ is the incident wave amplitude. Next, the analytic and numerical solution technique associated with the aforementioned boundary value problems will be discussed in two different subsequent Sections.

\section{Analytic method of solution}
\label{sec:3}
Using the eigenfunction expansion method, the velocity potential in each region is expanded in terms of appropriate eigenfunctions. The spatial velocity potentials $\phi_j$ for $j=1, 2$
satisfying Eq. (\ref{eq:1}) along with Eqs. (\ref{eq:2}), (\ref{eq:3}) and (\ref{eq:8}) are expressed as
\begin{eqnarray}
&&\phi_1(x,y)=-\frac{\mathrm{i}gH}{2\omega} \left(e^{\mathrm{i}\alpha_0 x} + A_0 e^{-\mathrm{i}\alpha_0 x}\right) I_0(y) + \sum_{n=1}^{\infty} A_n e^{\alpha_n x} I_n(y), \label{eq:9}\\
&&\phi_2(x,y)=-\frac{\mathrm{i}gH}{2\omega} B_0 e^{\mathrm{i}\alpha_0 x} I_0(y) + \sum_{n=1}^{\infty} B_n e^{-\alpha_n x} I_n(y), \label{eq:10}
\end{eqnarray}
where $A_0$, $B_0$, $A_n$ and $B_n$ for $n=1, 2, 3,...$ are the arbitrary complex constants to be determined and the eigenfunctions $I_0(y)$ and $I_n(y)$ are given by
\begin{equation}\label{eq:11}
I_0 (y)=\frac{\cosh\left(k_0(h-y)\right)}{\cosh k_0h},\quad I_n (y)=\frac{\cos \left(k_n(h-y)\right)}{\cos k_nh}\;\;\mbox{for}\;\; n=1, 2, 3, ...
\end{equation}
with $\alpha_n=\sqrt{\beta_0^2+k_n^2}$. It may be noted that the wave numbers $k_n$ satisfies $k_n \tan k_n h=-K$ for $n=1, 2, 3, ...$.
Applying the continuity of normal velocity as in Eq. (\ref{eq:7}) into Eqs. (\ref{eq:9})-(\ref{eq:10}), it is obtained that
\begin{equation}\label{eq:12}
1-A_0=B_0, \quad A_n=-B_n
\end{equation}
Using the relations (\ref{eq:12}) into Eqs. (\ref{eq:9}) and (\ref{eq:10}), and substituting the expressions of $\phi_1$ and $\phi_2$ into the right-hand side of the forcing term in Eq. (\ref{eq:4}), the complex amplitude of the
structural displacement $\chi(y)$ is expressed as
\begin{equation}\label{eq:13}
\chi(y)=A e^{\delta y}+B e^{-\delta y}+a_0A_0 I_0(y)+\sum_{n=1}^{\infty} a_n A_n I_n(y),
\end{equation}
where $\delta=\mathrm{i}\sqrt{B^\prime}$  with $B^\prime=\left(m_m \omega^2/\overline{T}\right)-\beta_0^2$. In Eq. (\ref{eq:13}), $A$ and $B$ are arbitrary constants need to be determined, and $a_0$ and $a_n$ are given by
\begin{equation}\label{eq:14}
a_0=-\frac{\rho g H}{\overline{T}} \frac{1}{\left(k_0^2 + B^\prime\right)}, \quad a_n=\frac{2 \mathrm{i} \rho \omega}{\overline{T}} \frac{1}{\left(k_n^2 - B^\prime\right)},
\end{equation}
Now, by using continuity of pressure as in Eq. (\ref{eq:7}) in Eq. (\ref{eq:5}), we get \begin{equation}\label{eq:15}
\mathrm{i}k_0 G_0 \left(\phi_1 - \phi_2\right) =
\left\{\begin{array}{l}
0 \qquad\qquad\quad \mbox{on}\; x=0,\; y \in L_g, \vspace{0.2cm}
\\
\displaystyle  \frac{\partial \phi_j}{\partial x}+\mathrm{i}\omega\chi \quad\; \mbox{on}\; x=0,\;y \in L_m.
\end{array}\right.
\end{equation}
Substituting for $\phi_1$, $\phi_2$, $\partial \phi_1/\partial x$ and $\chi$ from Eqs. (\ref{eq:9}), (\ref{eq:10}) and (\ref{eq:13}) into Eq. (\ref{eq:15}) and utilizing the relationships as in
(\ref{eq:12}), we obtain
\begin{equation}\label{eq:16}
\sum_{n=0}^{\infty} c_n A_n I_n =
\left\{\begin{array}{l}
0  \qquad\; \mbox{on}\; x=0,\; y \in L_g, \vspace{0.2cm}
\\
f(y) \quad\; \mbox{on}\; x=0,\;y \in L_m.
\end{array}\right.
\end{equation}
where
\begin{equation}\label{eq:17}
f(y)=\mathrm{i}\omega \left(Ae^{\delta y}+Be^{-\delta y}\right) + g_0 I_0 + \sum_{n=0}^{\infty} A_n b_n I_n,
\end{equation}
with
\begin{equation}\label{eq:18}
g_0=\frac{\alpha_0 g_0 H}{2\omega}, \;\; b_0=\mathrm{i}\omega a_0 - \frac{\alpha_0 g_0 H}{2\omega}, \;\; b_n=\mathrm{i}\omega a_n + \alpha_n,\;\; c_0=\frac{k_0 g H G_0}{\omega},\;\;\mbox{and}\;\;c_n=2\mathrm{i}k_0G_0.
\end{equation}
Using the orthogonal properties of $I_n (y)$ for $n=0, 1, 2,...$ and truncating the infinite series after $N$ terms, it is derived from Eqs. (\ref{eq:16}) - (\ref{eq:17}) that
\begin{eqnarray}
&&c_0 A_0 \langle {I_0, I_0} \rangle=\int_{L_m} I_0 f(y) dy, \label{eq:19}
\\
&&c_n A_n \langle {I_n, I_n} \rangle=\int_{L_m} I_n f(y) dy, (n=1, 2,...,N), \label{eq:20}
\end{eqnarray}
where $\displaystyle \langle {I_n, I_m} \rangle=\int_0^h I_n I_m dy$ with $\langle {I_n, I_m} \rangle=0,\;\;\mbox{for}\;\; n\neq m$. Now, Eqs. (\ref{eq:19}) and (\ref{eq:20})
give a system of $N+1$ equations. To determine the rest of the unknowns, the conditions prescribed at the membrane edges has to be used. Substituting the expression of $\chi(y)$
as in Eq. (\ref{eq:13}) into Eq. (\ref{eq:6}), we get the remaining two equations
\begin{eqnarray}
&&A e^{\delta a} + B e^{-\delta a} + a_0A_0 \frac{\cosh \left(k_0(h-a)\right)}{\cosh k_0h}+\sum_{n=0}^{N} a_n A_n \frac{\cos \left(k_n (h-a)\right)}{\cos k_nh}=0, \label{eq:21}
\\
&&A e^{\delta b} + B e^{-\delta b} + a_0A_0 \frac{\cosh \left(k_0(h-b)\right)}{\cosh k_0h}+\sum_{n=0}^{N} a_n A_n \frac{\cos \left(k_n (h-b)\right)}{\cos k_nh}=0.  \label{eq:22}
\end{eqnarray}
The system of equations (\ref{eq:19}) - (\ref{eq:22}) are solved to obtain the required unknowns.

\section{Numerical method of solution}
\label{sec:4}
In the present Section, a numerical solution based on boundary element method (see \cite{koley2014oblique}, \cite{behera2015wave} and \cite{koley2015interaction} for details) is developed for the boundary value problem as discussed in Sec. \ref{sec:2}. The schematic diagram for the computational domains used in boundary element method (BEM) is given in Fig. \ref{fig:1}b. Here, for the sake of clarity, various boundary conditions discussed in Sec. \ref{sec:2} are rewritten in the computational
boundaries for easy reference in the BEM. The free surface boundary condition as in Eq. (\ref{eq:2}) is rewritten as
\begin{equation}\label{eq:23}
\frac{\partial\phi_j}{\partial \emph{\textbf{n}}} + K \phi_j=0,~~\mbox{on}~~\Gamma_{fj},
\end{equation}
where $\partial/\partial \emph{\textbf{n}}$ denotes the normal derivative. The bottom boundary condition as in Eq. (\ref{eq:3}) is rewritten as
\begin{equation}\label{eq:24}
\displaystyle \frac{\partial \phi_j}{\partial \emph{\textbf{n}}}=0,~~\mbox{on}~~\Gamma_{bj}.
\end{equation}
Further, the far-field boundary condition as in Eq. (\ref{eq:8}) is rewritten as
\begin{equation}\label{eq:25}
\left\{\begin{array}{l}
\displaystyle \frac{\partial \left(\phi_1 - \phi_0\right)}{\partial \emph{\textbf{n}}} + \mathrm{i} \alpha_0 \left(\phi_1 - \phi_0\right) =0,~~\mbox{on}~~\Gamma_{c1}, \vspace{0.2cm}
\\
\displaystyle \frac{\partial \phi_2} {\partial \emph{\textbf{n}}} - \mathrm{i} \alpha_0 \phi_2=0,~~\mbox{on}~~\Gamma_{c2}.
\end{array}\right.
\end{equation}
It may be noted that theoretically far-field boundary boundary conditions are satisfied at $x\pm\infty$. However, for the sake of computation, the far-field boundaries $\Gamma_{cj}$ for $j=1, 2$ in Eq. (\ref{eq:25}) are taken three times water depth away from the membrane so that the effect of the local wave modes vanishes on $\Gamma_{cj}$.
The continuity of pressure and normal velocity at the gap regions $\Gamma_{gj}$ for $j=1, 2$ are given as
\begin{equation}\label{eq:26}
\left\{\begin{array}{l}
\displaystyle \phi_1=\phi_2,\qquad\mbox{on}~~\Gamma_{gj}  \vspace{0.2cm}
\\
\displaystyle \frac{\partial \phi_1} {\partial \emph{\textbf{n}}}=-\frac{\partial \phi_2} {\partial \emph{\textbf{n}}},~~\mbox{on}~~\Gamma_{gj}.
\end{array}\right.
\end{equation}
Moreover, the kinematic boundary condition on the membrane as in Eq. (\ref{eq:5}) is rewritten as
\begin{equation}\label{eq:27}
\displaystyle \frac{\partial \phi_1}{\partial \emph{\textbf{n}}}=- \frac{\partial \phi_2}{\partial \emph{\textbf{n}}}
=\mathrm{i}k_0G_0 \left(\phi_1-\phi_2\right) - \mathrm{i}\omega\chi,~~\mbox{on}~~\Gamma_{m}.
\end{equation}
Applying Green's integral theorem to the velocity potential $\phi \left(x, y\right)$ and the free space Green's function $G\left(x,y;x_0,y_0\right)$, an integral equation is obtained as
\begin{equation}\label{eq:28}
-\left(\begin{array}{c} \phi(x,y)
\\
\frac{1}{2}\phi(x,y)
\end{array}\right)
=\int_{\Gamma}\left(\phi\frac{\partial G} {\partial
\emph{\textbf{n}}}-G\frac{\partial\phi}{\partial \emph{\textbf{n}}}\right)\,
d\Gamma,~~\left(\begin{array}{c}
\mbox{if}\,(x,y)\,\in\Omega\,\mbox{but not on}~~ \Gamma
\\
\mbox{if}~~ (x,y)\,\mbox{on}~~ \Gamma
\end{array}\right).
\end{equation}
where the free space Green's function $G(x,y;x_0,y_0)$ is of the form (See \cite{koley2015oblique} for details)
\begin{equation}\label{eq:29}
G\left(x,\, y\,;\, x_{0},\, y_{0}\right)=-\frac{\mathrm{K_0}\left(\beta_0 r\right)} {2\pi},\qquad
r=\sqrt{\left(x-x_{0}\right)^{2}+\left(y-y_{0}\right)^{2}},
\end{equation}
where $\mathrm{K_0}$ is the  modified zeroth-order Bessel function of the second kind with $r$ being the distance from the field point $(x, y)$ to the source point $(x_{0}, y_{0})$.
The properties and asymptotic behavior of $\mathrm{K_0}\left(\beta_0 r\right)$ can be found in \cite{koley2014oblique}. Using the boundary conditions (\ref{eq:23}) - (\ref{eq:27}) into
Eq. (\ref{eq:28}), the integral equations in each fluid domain are obtained as
\begin{eqnarray}
\frac{1}{2}\phi_1&+&\int_{\Gamma_{f1}}\left(\frac{\partial G}{\partial \emph{\textbf{n}}}+KG \right) \phi_1 d\Gamma
+\int_{\Gamma_{g1}+\Gamma_{g2}}\left(\phi_{2}\frac{\partial G}{\partial \emph{\textbf{n}}}+ G\frac{\partial\phi_{2}}{\partial \emph{\textbf{n}}}\right) d\Gamma
+\int_{\Gamma_{m}}\left(\frac{\partial G}{\partial \emph{\textbf{n}}}- \mathrm{i}k_0 G_0 G \right) \phi_{1}d\Gamma
+\mathrm{i}k_0 G_0 \int_{\Gamma_{m}} G \phi_2 d\Gamma \nonumber\\
&+&\mathrm{i}\omega \int_{\Gamma_{m}}\chi G  d\Gamma
+\int_{\Gamma_{b1}}\phi_1 \frac{\partial G}{\partial \emph{\textbf{n}}} d\Gamma
+\int_{\Gamma_{l}}\left(\frac{\partial G}{\partial \emph{\textbf{n}}}+\mathrm{i}\alpha_0 G \right) \phi_1 d\Gamma
=\int_{\Gamma_{l}}\left(\frac{\partial \phi_0}{\partial \emph{\textbf{n}}}+\mathrm{i}\alpha_0 \phi_0 \right) G d\Gamma, \label{eq:30}
\\
\frac{1}{2}\phi_2&+&\int_{\Gamma_{f2}}\left(\frac{\partial G}{\partial \emph{\textbf{n}}}+K G \right) \phi_2 d\Gamma +\int_{\Gamma_{g1}+\Gamma_{g2}}\left(\phi_{2}\frac{\partial G}{\partial \emph{\textbf{n}}}- G\frac{\partial\phi_{2}}{\partial \emph{\textbf{n}}}\right) d\Gamma
+\int_{\Gamma_{m}}\left(\frac{\partial G}{\partial \emph{\textbf{n}}}- \mathrm{i}k_0 G_0 G \right) \phi_{2}d\Gamma
+\mathrm{i}k_0 G_0 \int_{\Gamma_{m}} G \phi_1 d\Gamma \nonumber\\
&-&\mathrm{i}\omega \int_{\Gamma_{m}}\chi G d\Gamma
+\int_{\Gamma_{b2}}\phi_2 \frac{\partial G}{\partial \emph{\textbf{n}}} d\Gamma
+\int_{\Gamma_{r}}\left(\frac{\partial G}{\partial \emph{\textbf{n}}}-\mathrm{i}\alpha_0 G \right) \phi_2 d\Gamma
=0,\label{eq:31}
\end{eqnarray}
In Eqs. (\ref{eq:30}) and (\ref{eq:31}), all the boundary conditions of $\phi_1$ and $\phi_2$ except the dynamic membrane boundary condition as in Eq. (\ref{eq:4}) has been used. Since the body boundary condition is not known in advance in contrast to the rigid body hydrodynamics, Eqs. (\ref{eq:30}) and (\ref{eq:31}) cannot be solved without coupling with the membrane
equation of motion given by Eq. (\ref{eq:4}). To solve Eqs. (\ref{eq:30}) and (\ref{eq:31}), the entire boundary is discretized into a finite number of segments, called boundary
elements. On each boundary element, the velocity potential and its normal derivative are assumed to be constants and the influence coefficients $\int G$ and
$\int \partial G/\partial \emph{\textbf{n}}$ are evaluated numerically using Gauss-Legendre quadrature formula (see \cite{au1982numerical} and \cite{koleythesis} for details).
Utilizing Eqs. (\ref{eq:4}), the discrete form of the membrane equation of motion for the $j^{\mathrm{th}}$ element is given by (see \cite{kim1996flexible})
\begin{equation}\label{eq:32}
\frac{\mathrm{i}\rho\omega}{\overline{T}} \left(\phi_{1j}-\phi_{2j}\right) \Delta_j - \frac{\left(\chi_j-\chi_{j-1}\right)}{\Delta_j^m} + \frac{\left(\chi_{j+1}-\chi_{j}\right)}{\Delta_{j+1}^m}
=-\left(\frac{m_m \omega^2}{\overline{T}} - \beta^2_0\right) \Delta_j \chi_j,
\end{equation}
where $\Delta_j$ is the length of the $j^{\mathrm{th}}$ segment and $\Delta_j^m=\left(\Delta_j+\Delta_{j+1}\right)/2$. Using this kind of discrete model,
membrane edge conditions can be easily implemented. So, the discretized form of the integral equations (\ref{eq:30}) and (\ref{eq:31}) together with (\ref{eq:32}) give a system
of $\left(N+N_m-2\right)$ number of equations consist of $\phi$ and $\partial \phi/\partial \emph{\textbf{n}}$ and $\chi$ with $N$ and $N_m$ being the total number of boundary elements on the
total boundaries of regions 1 and 2 of the physical domain, and on the membrane surface respectively. It may be noted that, the
common boundaries of regions $1$ and $2$ are considered only one time to count $N$ and in the two boundary elements near to the edges of the membrane, the edges condition as in Eq. (\ref{eq:6}) is used.  These system of $\left(N+N_m-2\right)$ linear algebraic equations are finally solved to get the required unknowns. The convergence criteria of the numerical solutions and related discussions are given in \cite{kim1996flexible}, and the details are deferred here.

\section{Results and Discussions}
\label{sec:5}
In this Section, a MATLAB program is developed to analyze the effects of different wave and
structural parameters on wave scattering. In all the figures, lines and symbols correspond to
the analytic and numerical solutions respectively. In the current study, the values of the parameters $G_0=0.25+0.25\mathrm{i}$, $m_1=m_m/(\rho h)=0.01$,
$T_1=\overline{T}/\left(\rho g h^2\right)=0.4$, $\theta=30^{\circ}$, $H=h/10$ are kept fixed unless it is mentioned.
The reflection and transmission coefficients are computed using the formulae
\begin{equation}
K_r=\left|\frac{2\omega R}{gH}\right|,\quad K_t=\left|\frac{2\omega T}{gH}\right|
\end{equation}
The non-dimensional horizontal wave force acting on the membrane $K_f$ is given by the formula
\begin{equation}
K_f=\frac{\omega}{g h^2}\left| \int_{L_m} \left[\phi_1(0, y) - \phi_2(0, y)\right]\right|.
\end{equation}
The non-dimensional membrane deflection $\varsigma$ is given by the formula
\begin{equation}
\varsigma=\left|\frac{\chi}{h}\right|.
\end{equation}
Hereafter, several numerical results have been plotted and analyzed to show the effectiveness of a permeable flexible membrane as a wave barrier.

\begin{figure}[ht!]
\begin{center}
\subfigure[]{\label{fig:3a}\includegraphics[width=8.4cm,height=6cm]{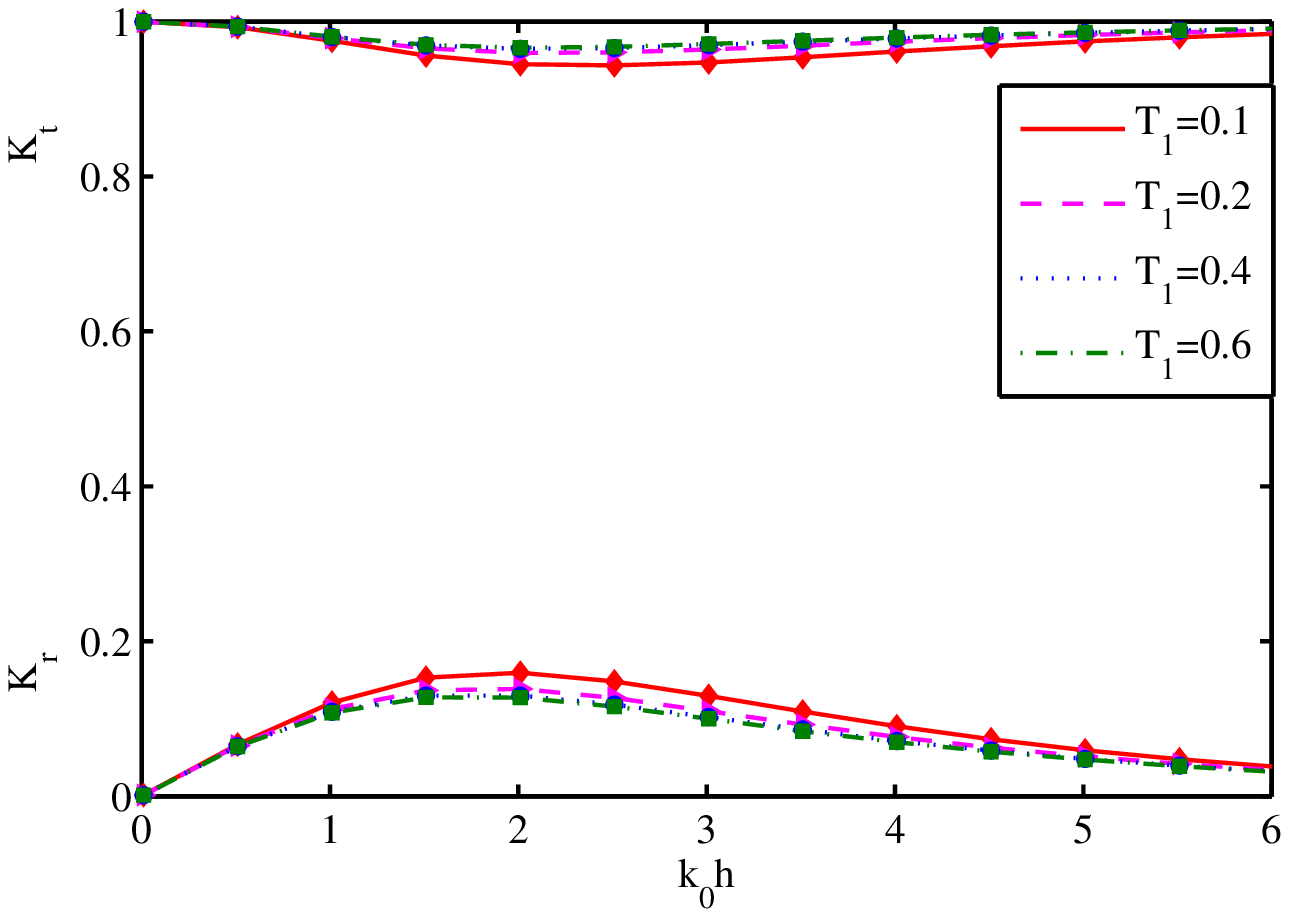}}
\subfigure[]{\label{fig:3b}\includegraphics[width=8.4cm,height=6cm]{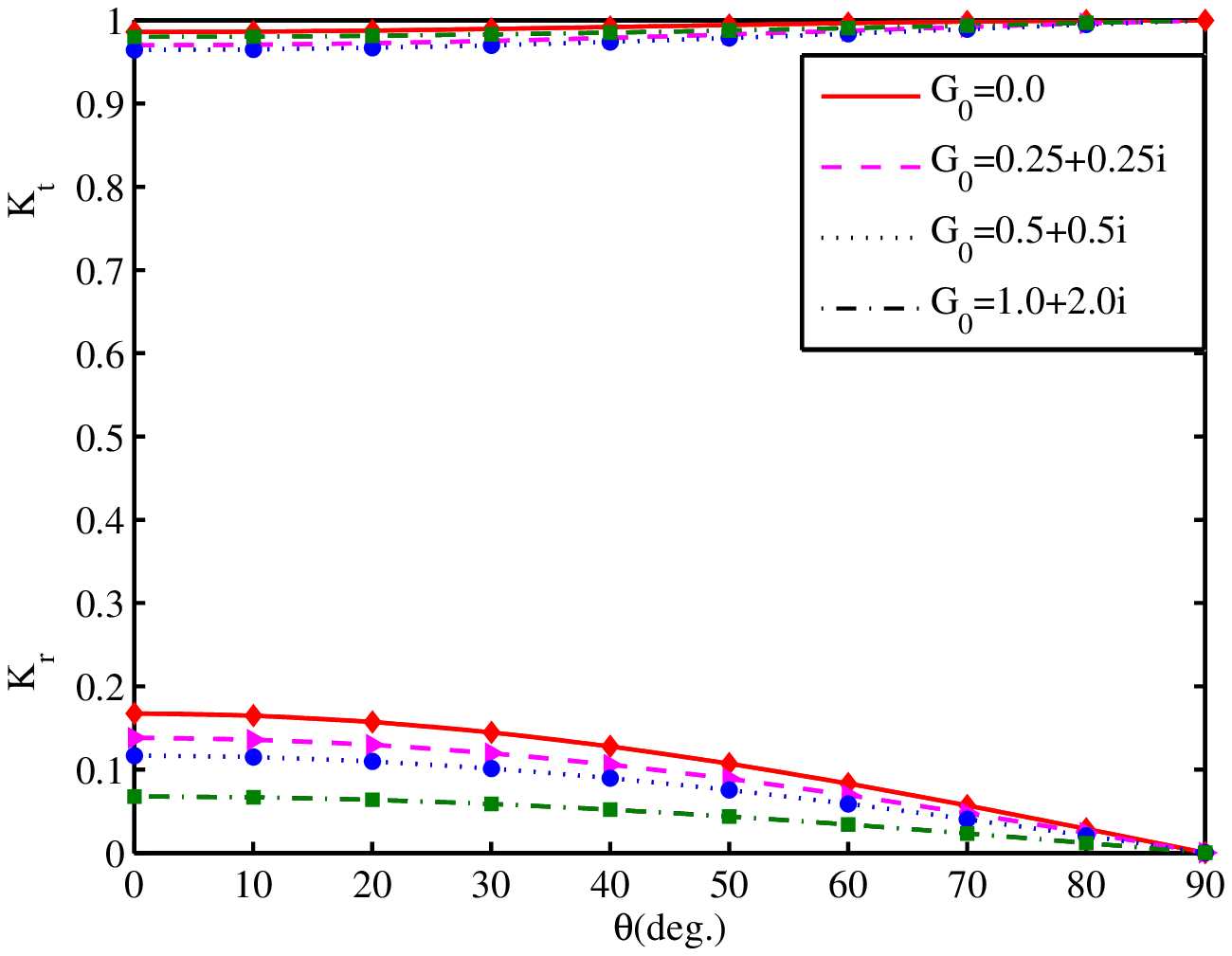}}\vspace{-0.5cm}
\end{center}
\caption{Variation of $K_r$ and $K_t$ vs. (a) $k_0h$ for different values of $T_1$ with $G_0=0.25+0.25\mathrm{i}$, and (b) $\theta$ for different values of
$G_0$ with $T_1=0.4$, $a/h=0.2$, $b/h=0.8$.}
\label{fig:2}
\end{figure}
In Figs. \ref{fig:2}(a) and (b), the reflection and transmission coefficients are plotted as a function of non-dimensional wave number $k_0h$ and angle of incidence $\theta$
for various values of normalized membrane tension $T_1$ and porous-effect parameter $G_0$ respectively for a submerged membrane. From Fig. \ref{fig:2}(a), it is observed that with an increase in membrane tension, reflection coefficient decreases and reverse pattern is observed for transmission coefficient. In case of
Fig. \ref{fig:2}(b), it is seen that with an increase in the angle of incidence $\theta$, reflection coefficient $K_r$ decreases and reverse pattern is observed for transmission coefficient $K_t$. Moreover, for a particular value of $\theta$, with an increase in porosity of the membrane, wave reflection decreases. This may be due to the reason that with an increase in porosity of the structure, more wave energy is dissipated by the structure.

\begin{figure}[ht!]
\begin{center}
\subfigure[]{\label{fig:3a}\includegraphics[width=8.4cm,height=6cm]{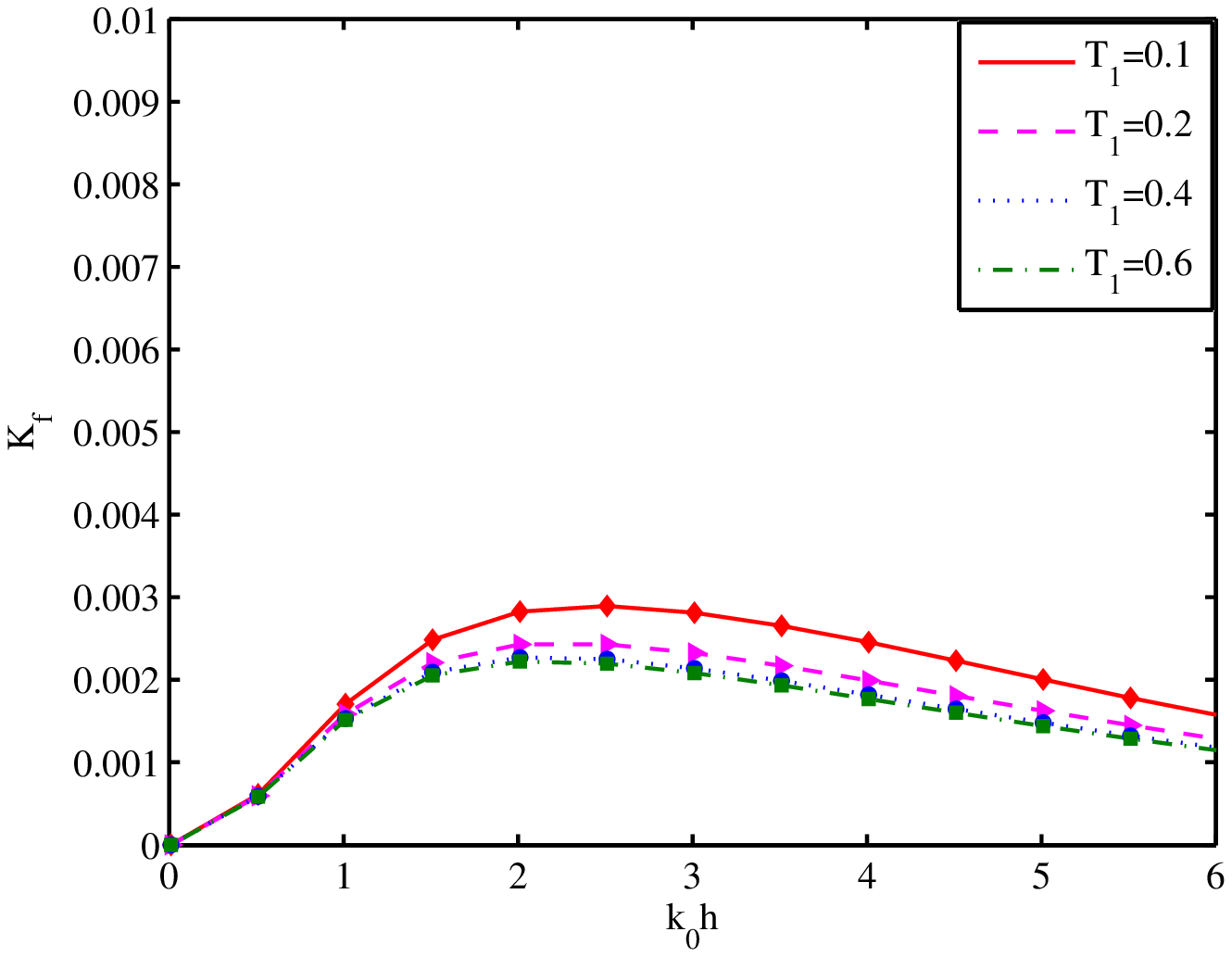}}
\subfigure[]{\label{fig:3b}\includegraphics[width=8.4cm,height=6cm]{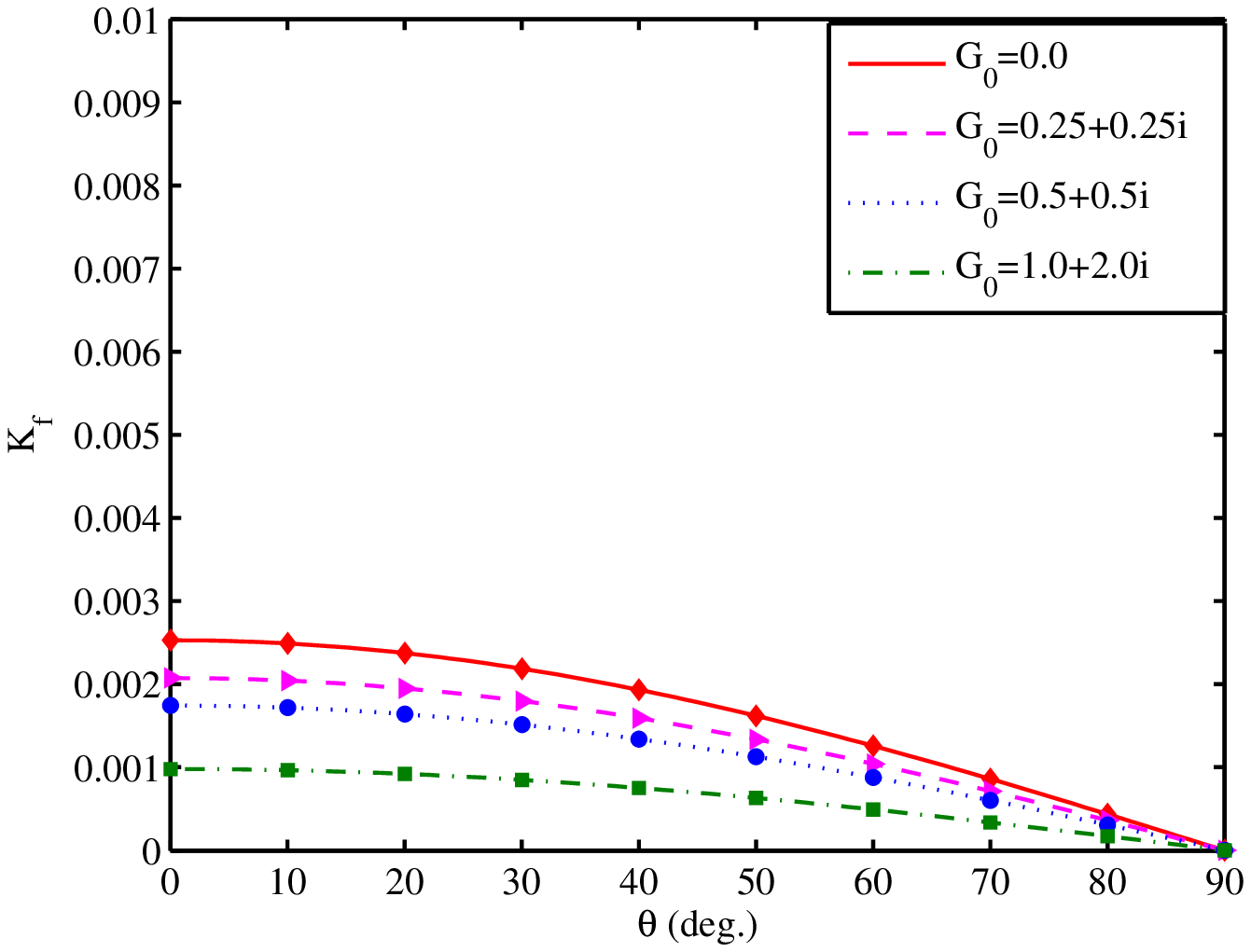}}\vspace{-0.5cm}
\end{center}
\caption{Variation of $K_f$ vs. (a) $k_0h$ for different values of $T_1$ with $G_0=0.25+0.25\mathrm{i}$, and (b) $\theta$ for different values of
$G_0$ with $T_1=0.4$, $a/h=0.2$, $b/h=0.8$.}
\label{fig:3}
\end{figure}
In Figs. \ref{fig:3}(a) and (b), horizontal wave force $K_f$ acting on a submerged
membrane versus (a) non-dimensional wave number $k_0h$ is plotted for different
values of membrane tension $T_1$, and (b) angle of incidence $\theta$ for different
values of porous-effect parameter $G_0$. The general pattern of the wave forces
$K_f$ are similar to that of reflection coefficient $K_r$ as in Fig. \ref{fig:2}. By
comparing Fig. \ref{fig:3} with that of Fig. \ref{fig:2}, it can be concluded that
optima in wave reflection corresponds to that of wave force acting on the membrane.

\begin{figure}[ht!]
\begin{center}
\subfigure[]{\label{fig:3a}\includegraphics[width=8.4cm,height=6cm]{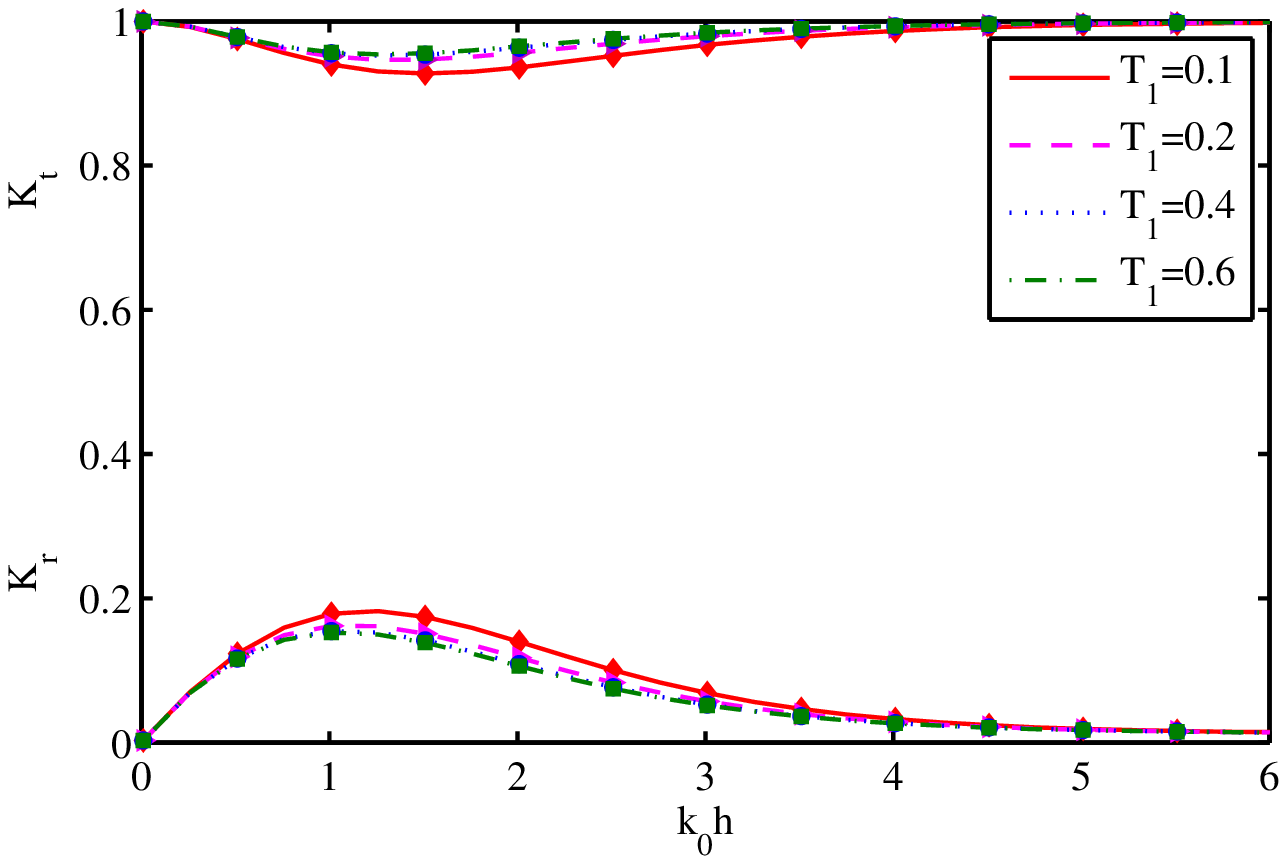}}
\subfigure[]{\label{fig:3b}\includegraphics[width=8.4cm,height=6cm]{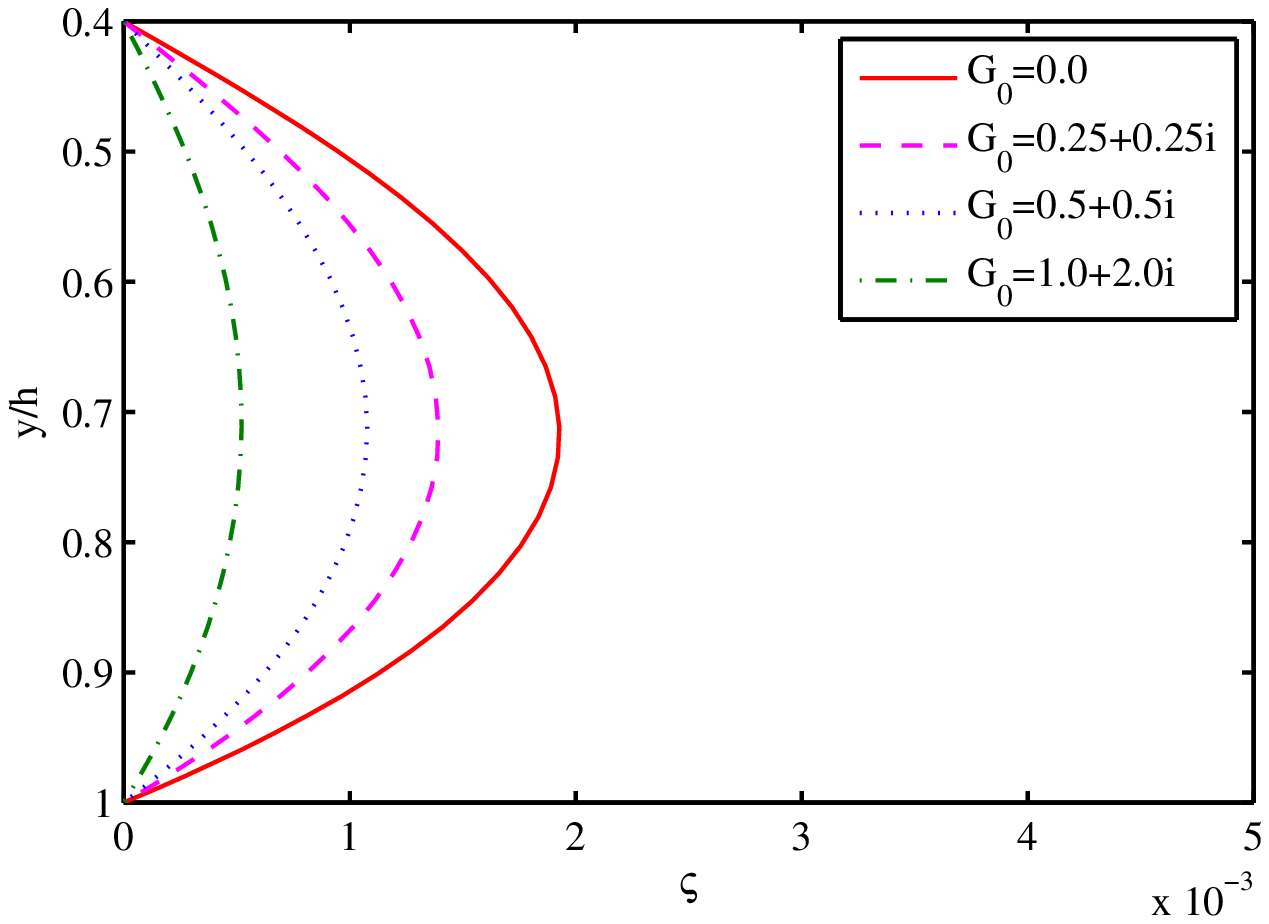}}\vspace{-0.5cm}
\end{center}
\caption{Variation of (a) $K_r$ and $K_t$ vs. $k_0h$ for different values of $T_1$ with $G_0=0.25+0.25\mathrm{i}$, and (b) membrane
deflection $\varsigma$ for different values of  $G_0$ with $T_1=0.4$, $a/h=0.4$, $b/h=1.0$.}
\label{fig:4}
\end{figure}
In Fig. \ref{fig:4}(a), the reflection and transmission coefficients $K_r$ and $K_t$
versus non-dimensional wave number $k_0h$ are plotted for various values
of the membrane tension $T_1$ for a bottom-standing membrane barrier.
The general pattern of $K_r$ and $K_t$ are similar to that of Fig. \ref{fig:2}(a) in case
of submerged membrane. However, a comparison between Figs. \ref{fig:2}(a) and \ref{fig:4}(a) reveals that maxima in the wave reflection and correspondingly minima in the
wave transmission occurs for smaller values of $k_0h$ in case of bottom-standing membrane compared to that of submerged membrane. On the other hand, Fig. \ref{fig:4}(b)
shows the displacement profiles for a bottom-standing membrane for various values
of porous-effect parameter $G_0$. As the membrane is fixed at both ends, there are no deflection of the membrane near both the ends and maximum deflection occurs near the center of the membrane. It is also observed from Fig. \ref{fig:4}(b) that the deflection of the
membrane decreases as the absolute value of the porous-effect parameter $G_0$ increases.
This may be due to the fact that with an increase in the structural porosity, wave impact on the membrane reduces which in turn reduces the deflection of the membrane.

 \begin{figure}[ht!]
 \begin{center}
 \subfigure[]{\label{fig:3a}\includegraphics[width=8.4cm,height=6cm]{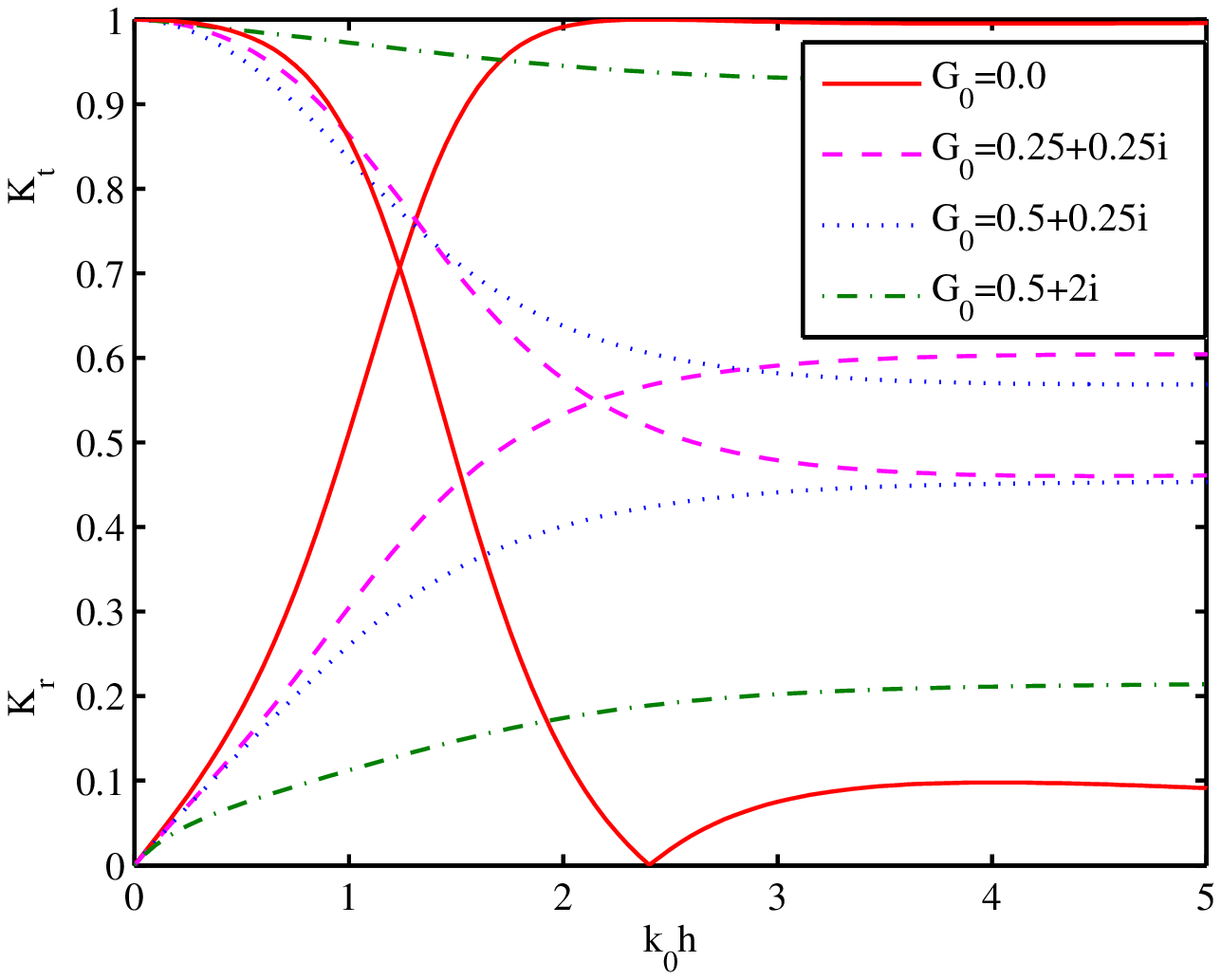}}
 \subfigure[]{\label{fig:3b}\includegraphics[width=8.4cm,height=6cm]{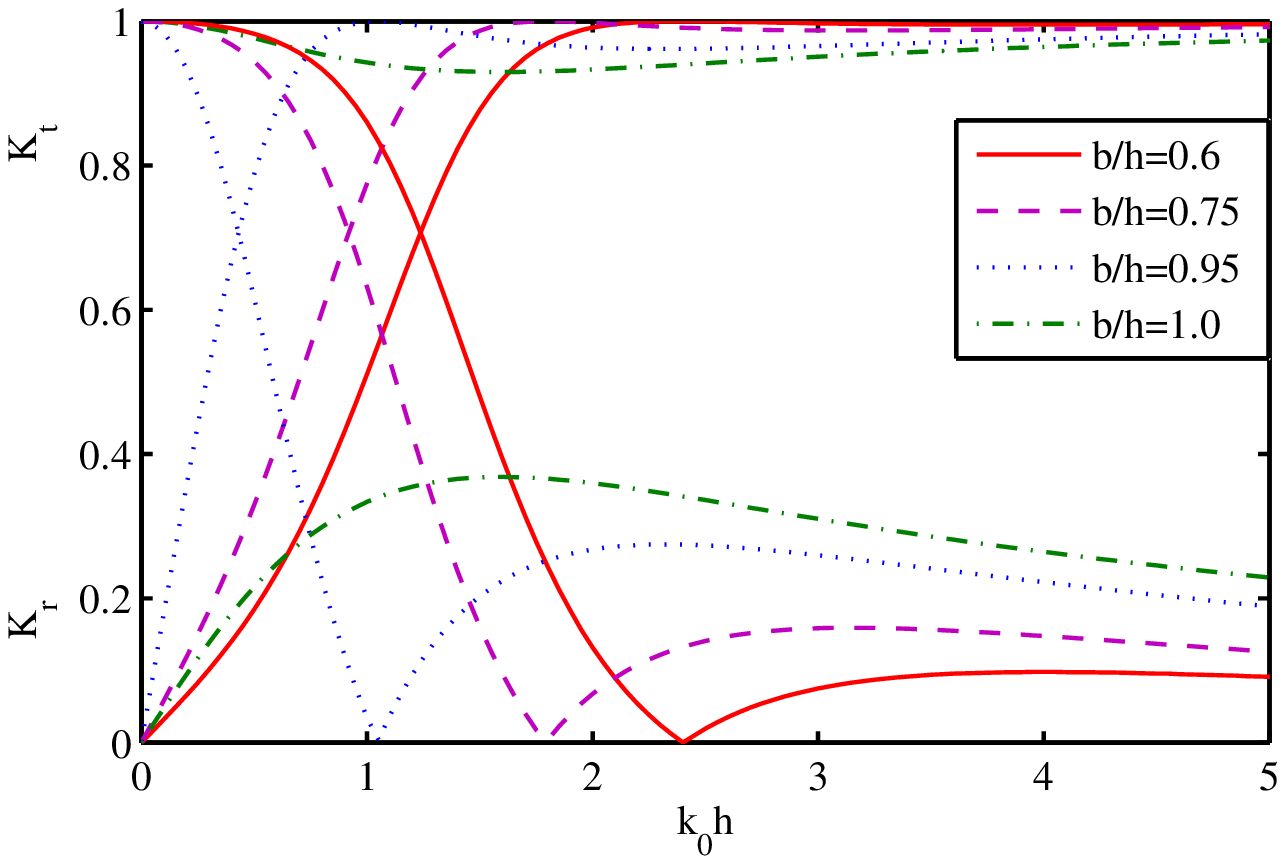}}\vspace{-0.5cm}
 \end{center}
 \caption{Variation of  $K_r$ and $K_t$ vs. $k_0h$ for different values of (a) $G_0$ with
 $a/h=0.0$, $b/h=0.6$, and (b) $b/h$ with $G_0=0.0$, $T_1=0.4$, $a/h=0.0$.}
 \label{fig:5}
 \end{figure}
 In Fig. \ref{fig:5}, the reflection and transmission coefficients $K_r$ and $K_t$
 versus non-dimensional wave number $k_0h$ are plotted for various values of the porous-effect parameter $G_0$, and (b) membrane length $b/h$ for a surface-piercing
 impermeable membrane barrier. A comparison with Figs. \ref{fig:2}(a) and \ref{fig:4}(a)
 demonstrates that more wave reflection and less transmission occurs in case of surface-piercing permeable membrane barrier compared to that of the bottom-standing and submerged membrane barrier. On the other hand, Fig. \ref{fig:5}(b) reveals that near $k_0h \approx 0$, full wave reflection and zero transmission occurs in case of complete impermeable membrane barrier. However, reverse pattern is observed in case of partial impermeable membrane barrier
 irrespective of membrane length $b/h$. This phenomena occurs due to the occurrence of wave diffraction in the gap region below the surface-piercing membrane wave barrier

\section{Conclusions}
\label{sec:6}
The interaction of obliquely incident waves with vertical flexible porous membranes of
different configurations are investigated using the linear water wave theory and small
amplitude structural responses. Solutions for different types of barrier configurations
are obtained using eigenfunction expansion-based analytic and BEM-based numerical method. It is observed that results
obtained by the two methods are in good agreement. The effectiveness of the surface-piercing
membrane barrier over the bottom-standing as well as submerged membrane are
analyzed from the reflection and transmission coefficients. The inclusion of structural
porosity to the membrane wave barrier significantly reduces the wave forces acting on the
barrier as well as the deflection of the barrier. With appropriate choice of tensile force and
porous-effect parameter, a permeable flexible membrane can act as an effective wave barrier.

%\section*{References}
%\bibliographystyle{elsarticle-harv}

\end{document}